\documentclass[%
 reprint,
 amsmath,amssymb,
 aps,
superscriptaddress]{revtex4-2}

\usepackage{graphicx}
\usepackage{amsmath}
\usepackage{verbatim}
\usepackage{subcaption}
\usepackage{physics}
\usepackage{xcolor}
\usepackage{float}  
\usepackage{hyperref}
\usepackage{dcolumn}
\usepackage{bm}
\begin{document}

\preprint{APS/123-QED}

\title{A comparison of different master equations for driven-dissipative dynamics in composite quantum systems: Dispersive readout in structured electromagnetic environments}

\author{Prakritish Gogoi}
\affiliation{Sorbonne Université, Institut des NanoSciences de Paris, 4 place Jussieu, 75005 Paris, France}
\author{Angela Riva}
\affiliation{LPENS, Département de Physique, École Normale Supérieure, Centre Automatique et Systèmes (CAS), MINES ParisTech, Université PSL, Sorbonne Université, CNRS, Inria, 75005 Paris, France}
\author{{\'E}mile Cochin}
\affiliation{Sorbonne Université, Institut des NanoSciences de Paris, 4 place Jussieu, 75005 Paris, France}
\author{Alex Chin}
\affiliation{Sorbonne Université, Institut des NanoSciences de Paris, 4 place Jussieu, 75005 Paris, France}
\affiliation{CNRS, Institut des NanoSciences de Paris, 4 place Jussieu, 75005 Paris, France}

\date{\today}

\begin{abstract}
Driven–dissipative qubit–resonator dynamics, which are the basis of most dispersive superconducting qubit measurement schemes, are often modeled with Lindblad master equations built from subsystem-local jump operators, even when the qubit and resonator are appreciably hybridized. In this work we revisit this setting using a microscopic Bloch–Redfield approach, where dissipation is constructed in the eigenbasis of the coupled qubit–resonator Hamiltonian with a complete, frequency dependent,  open system description of the transmission-line `environment'. Here, we show that the Lindblad and the Bloch-Redfield decay rates can be quantitatively different in the absence of driving, while in the driven case we demonstrate that the time-independent Redfield dissipator and its time-dependent generalization, can show qualitatively different behaviors, as a function of driving strength. Finally, we investigate the effects of driving in structured spectral densities, recovering the suppression of measurement-induced relaxation in the presence of a so-called Purcell filter. 
\end{abstract}

\maketitle

\section{\label{sec:level1}Introduction}
Driven-dissipative dynamics are ubiquitous in controlled quantum systems and constitute one of the major forefronts of present day quantum research. In particular, there are a wide variety of use cases that demonstrate how understanding driven-dissipative systems could unlock the potential and reliability of future quantum technologies  \cite{aspelmeyer_optomech_2014, gardiner_inputoutput_1985, ritsch_coldatoms_2013, carusotto_quantumfluids_2013}. However, they are, in general, hard problems to solve due to their complexity, and this is even more onerous when we are dealing with systems which consist of distinct components that are strongly coupled to each other. The dispersive qubit-resonator model \cite{bianchetti_dispersivereadout_2009} in circuit QED is an important example of how coupling between two systems can result in rich driven-dissipative dynamics. It is essentially a qubit readout technique in which the qubit is coupled to a far-detuned cavity whose frequency then depends on the state of the embedded qubit. The shift, and therefore the state of the qubit, can then be inferred by measuring how these shifts modify the cavity's response to external microwave drives (see Fig.\ref{fig_driven:subfig1}). 

This measurement scheme is ideally quantum nondemolition (QND) \cite{imoto_qndphoton_1989}, meaning that increasing the drive amplitude should enable faster, higher-fidelity qubit state discrimination. Measured readout fidelity instead saturates and decreases beyond a certain drive amplitude, indicating a departure from ideal QND operation \cite{walter_rapidreadout_2017, sank_measurementtransition_2016, dai_drivetransitions_2026}. One manifestation of this breakdown is a drive-dependent change of the qubit energy relaxation time $T_1$.

While microscopic derivations of effective master equations for multi-level weakly anharmonic systems, which account in principle for the frequency dependence of the electromagnetic environment, predict the opposite trend \cite{petrescu_readout_2020, malekakhlagh_lifetimerenorm_2020}, the standard approach for predicting these dynamics is to use a Lindblad master equation that models the dissipation phenomenologically with jump operators that capture the single-photon loss processes into the transmission line \cite{sete_purcell_2014, blais_cavity_2004}. While doing so, it is implicitly assumed that the qubit and resonator are independent subsystems, with dissipation acting locally on the resonator as if the qubit-resonator hybridization plays no role in mediating energy loss to the environment. This, however, is a strong assumption and can lead to inconsistencies that are exquisitely sensitive to the operating parameter regime \cite{beaudoin_dissipation_2011}. Hence, for coupled systems, both in the present example, and more widely,  it is important to understand how different microscopic assumptions used in modelling dissipation can lead to both quantitatively different predictions and, as we shall show, even \emph{qualitatively} different responses under driven conditions.

In this work we address these limitations using the Bloch-Redfield master equation \cite{jeske_bloch-redfield_2015, blum_density_2012}, which constructs the dissipator in the eigenbasis of the coupled qubit-resonator Hamiltonian and incorporates the frequency dependence of the environment through its spectral density. This yields frequency-selective transition rates that are absent in the standard Lindblad treatment. We consider both a static dissipator, built from the undriven eigenbasis, and a time-dependent generalization that accounts for the dressing of the system by the drive. Throughout, we focus on relaxation induced by the coupling to the transmission line, setting aside pure dephasing noise and its associated dressed dephasing effect \cite{boissonneault_dresseddephasing_2008, boissonneault_dispersive_2009, slichter_qubitmixing_2012, gambetta_qtraj_2008}, in order to isolate the intrinsic measurement-induced relaxation mechanisms.

Depending on the type of treatment of the dissipation (time-dependent or independent), we find that the rotating-wave approximation applied to the drive Hamiltonian produces unphysical relaxation trends within the time-dependent formulation, while retaining the full cosine drive restores physically consistent behavior. We further demonstrate that structured electromagnetic environments, such as Purcell filters relevant to realistic circuit impedances \cite{hutin_cavityenergy_2024, sunada_intrinsicpurcell_2022}, can be incorporated directly through the spectral density function — at significantly lower computational cost than numerically exact methods \cite{strathearn_tempo_2018, lacroix_mpsdynamics_2024, cochin_processtensors_2026, riva_dispersivereadout_2026}. 

The article is organized in the following way. In Sec.II we introduce the model that describes our system and the environment. In Sec.III we discuss the methods namely the Lindblad and the Bloch-Redfield model used to tackle the problem. Then we discuss both the undriven and driven case with results in Sec.IV A and Sec.IV B respectively. At the end in Sec.IV B we discuss the Purcell filter using the Bloch-Redfield formalism. 
\section{\label{sec:level2}The Model}
The Hamiltonian of the coupled system of the qubit which is considered to be a two-level system and the resonator which is a harmonic oscillator driven by a microwave drive can be written as
\begin{subequations}{\label{modeleq}}
    \begin{align}
        \hat{H}_S(t) &= \hat{H}_0 + \hat{H}_d(t),\label{modeleq: a}\\
        \hat{H}_0 &= \frac{\omega_q}{2}\hat{\sigma}_z + \omega_r \hat{a}^\dagger\hat{a} + g\hat{\sigma}_x(\hat{a} + \hat{a}^\dagger),\label{modeleq: b}\\
        \hat{H}_d(t) &= \eta\cos(\omega_dt)(\hat{a} + \hat{a}^\dagger),
    \end{align}
\end{subequations}
where $\omega_q$, $\omega_r$, $\omega_d$ and $g$ are the qubit frequency, resonator frequency, drive frequency and the qubit-resonator coupling strength respectively. $\hat{\sigma}_z$, $\hat{\sigma}_x$ are the Pauli X and Z operators and $\hat{a}$, $\hat{a}^\dagger$ are the creation and annihilation operators of the harmonic oscillator. Here, $\hat{H}_0$ is the Rabi Hamiltonian with $\hat{H}_d(t)$ as the microwave drive that is sent into the cavity via the transmission line for qubit state readout. If we impose a rotating wave approximation in the qubit-resonator coupling term in (\ref{modeleq: b}), we end up with the Jaynes-Cummings Hamiltonian which is given by
\begin{equation}
    \hat{H}_{JC} = \frac{\omega_q}{2}\hat{\sigma}_z + \omega_r\hat{a}^\dagger\hat{a} + g(\hat{\sigma}_-\hat{a}^\dagger + \hat{\sigma}_+\hat{a}),
    \label{eqjc}
\end{equation}
although we are going to work with the Rabi Hamiltonian $\hat{H}_0$ throughout the entire work unless mentioned and we have assumed $\hbar = 1$. The full Hamiltonian along with the environment can be written as 
\begin{equation}
    \hat{H} = \hat{H}_S(t) + \hat{H}_E + \hat{H}_I,
    \label{tHamiltonian}
\end{equation}
where $\hat{H}_E$ is the Hamiltonian of the environment which, in our case is the semi-infinite coplanar waveguide transmission line to which the single mode cavity is capacitively coupled. The Hamiltonian of the transmission line in the continuum limit can be written as \cite{blais_circuit_2021}
\begin{equation}
    \hat{H}_E = \int_0^\infty d\omega \omega\hat{b}_\omega^\dagger\hat{b}_\omega,
    \label{envHamil}
\end{equation}
where the operators $\hat{b}_\omega$ and $\hat{b}^\dagger_\omega$ satisfy the commutation relation $[\hat{b}_\omega, \hat{b}_\omega^\dagger] = \delta(\omega - \omega')$. 
The interaction Hamiltonian $\hat{H}_I$ between the cavity and the semi-infinite transmission line is given by \cite{blais_circuit_2021}
\begin{equation}
     \hat{H}_I = (\hat{a} + \hat{a}^\dagger)\int_0^\infty d\omega\sqrt{J(\omega)}(\hat{b}_\omega + \hat{b}_\omega^\dagger).
     \label{IntHamil3}
\end{equation}
Here, $J(\omega)$ is the commonly used Ohmic spectral density of the transmission line \cite{yurke_input-output_2004, cattaneo_resistiveelements_2021}, which takes the form
\begin{equation}
    J(\omega) = \frac{\pi}{2}\alpha\omega \Theta(\omega_c - \omega)
    \label{sdActual}
\end{equation}
where $\alpha$ is the dimensionless coupling constant and $\omega_c$ is the cutoff frequency which we have considered in our case to be $\omega_c = 2\omega_r$ so that it encompasses all the relevant transition frequencies.
\begin{figure}
    \centering
    \includegraphics[width=0.8\columnwidth]{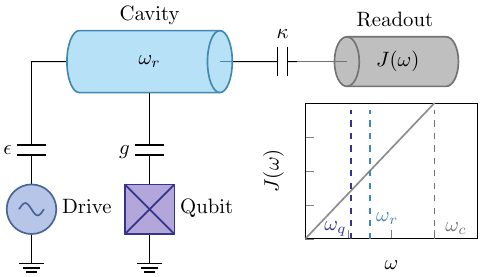}
    \caption{Schematic of the dispersive readout circuit, also showing the spectral density $J(\omega)$ of the transmission-line environment with the typical frequency scales of the qubit ($\omega_q$), cavity ($\omega_r$) and environmental cut-off frequency ($\omega_c$) indicated by the vertical lines.}
    \label{fig:placeholder}
\end{figure}
\section{Methods}

\subsection{\label{sec:level2}Lindblad equation}
The standard approach to simulate the dynamics of the driven dissipative qubit-resonator Hamiltonian is to use a Lindblad master equation where the time evolution of the density matrix is given by \cite{sete_purcell_2014}
\begin{subequations}\label{eq:Lindblad}
\begin{align}
  \dot{\hat{\rho}} &= -i[\hat{H}_S(t), \hat{\rho}] + \kappa\mathcal{D}[\hat{a}]\hat{\rho},\;,\label{eq:La}\\
\kappa\mathcal{D}[\hat{a}]\hat{\rho} &= \kappa\left(\hat{a}\hat{\rho}\hat{a}^\dagger - \frac{1}{2}\{\hat{a}^\dagger \hat{a}, \hat{\rho}\}\right),\label{eq:Lb}
\end{align}
\end{subequations}
that models the dissipation with the single photon loss dissipator $\mathcal{D}[\hat{a}]$. Here, $\kappa$ is the single photon loss rate which is related to the quality factor of the cavity as $\kappa = \omega_c/Q$. While this looks reasonable at a first glance, we will later see that constructing the dissipator with operators local to a subsystem can cause issues which lead to inaccurate dynamics.
The Lindblad master equation also neglects the frequency dependence of coupling to the environment, as the dissipator $\mathcal{D}[\hat{a}]$ is merely scaled by the single-photon loss rate $\kappa$. On expressing $\mathcal{D}[\hat{a}]$ in the eigenbasis of the qubit-resonator Hamiltonian $\hat{H}_0$, it can be shown that the Lindblad rates are given by
\begin{align}
\Gamma_L
&= \kappa\left|\bra{\mu}\hat{a}\ket{\nu}\right|^2,
\label{LindRates}
\end{align}
where $\{\ket{\mu}\}$ are the eigenstates of $\hat{H}_0$. We see that all of the transitions $\mu \leftarrow \nu$ have equal spectral weight $\kappa$. In other words, the Lindblad model sees the environment as a frequency independent flat noise source. This assumption can be problematic in some scenarios, considering that the spectrum of the transmission line is actually frequency dependent \cite{yurke_input-output_2004, cattaneo_resistiveelements_2021},  and also, if we want to include features like a Purcell filter in the transmission line (see Section \ref{purcell_filt_sec}).
\subsection{Bloch-Redfield Equation}
The first part of this section discusses the construction of the time-independent dissipator i.e. the Redfield tensor, built from the eigenbasis of the bare system Hamiltonian $\hat{H}_0$. Then we move on to building the dissipator in the instantaneous eigenbasis of the driven Hamiltonian $\hat{H}_S(t)$ thus rendering the dissipation time-dependent. The idea behind using two different approaches to construct the dissipator is to check if the inclusion or exclusion of counter-rotating terms in the drive has any influence on the long term behavior of the dynamics. We use the Bloch-Redfield implementation found in the QuTip 5.2 package \cite{lambert_qutip_2026} for our numerical simulations.
\subsubsection{Static Redfield tensor}
We start from the microscopic picture with the von Neumann equation in the interaction picture with an interaction of the type $\hat{H}_I =\hat{A} \otimes \hat{B}$ where $\hat{A}$ and $\hat{B}$ are operators that act on the system and the environment Hilbert space respectively. Following the procedures mentioned in \cite{breuer_theory_2007}, in the Redfield equation we perform the Born-Markov approximations and trace out the environment to get the matrix form expressed in the eigenbasis of $\hat{H}_0$ as
\begin{equation}
    \dot{\rho}_{\mu\nu}(t) = -i\omega_{\mu\nu}\rho_{\mu\nu}(t) + \sum_{\mu',\nu'}\mathcal{R}_{\mu\nu\mu'\nu'}\rho_{\mu'\nu' }(t)
    \label{BRmaster}
\end{equation}
where the static Redfield tensor $\mathcal{R}_{\mu\nu\mu'\nu'}$ is given by \cite{blum_density_2012}
\begin{equation}\label{eq:RedfieldBlum}
\begin{aligned}
\mathcal{R}_{\mu\nu\mu'\nu'}
&= \Gamma^{(+)}_{\nu'\nu\mu\mu'} + \Gamma^{(-)}_{\nu'\nu\mu\mu'} \\
&\quad {}- \delta_{\nu'\nu}\sum_k \Gamma^{(+)}_{\mu k k \mu'}
      - \delta_{\mu'\mu}\sum_k \Gamma^{(-)}_{\nu' k k \nu}.
\end{aligned}
\end{equation}
with
\addtocounter{equation}{-1}
\begin{subequations}\label{eq:Rsubs}
\begin{align}
\Gamma^{(+)}_{\nu'\nu\mu\mu'} &= J(\omega_{\mu'\mu})A_{\nu'\nu}A_{\mu\mu'}, \label{eq:Rsubs:a}\\
\Gamma^{(-)}_{\nu'\nu\mu\mu'} &= J(\omega_{\nu'\nu})A_{\nu'\nu}A_{\mu\mu'}. \label{eq:Rsubs:b}
\end{align}
\end{subequations}
$J(\omega)$ is the spectral density of the environment which is related to the Fourier transform of the environment correlation functions  as $J(\omega) = \int_{-\infty}^\infty C(\tau)e^{i\omega\tau}d\tau$ and $C(\tau) = Tr_E[\hat{B}(t)\hat{B}(t - \tau)] = \langle \hat{B}(\tau)\hat{B}(0)\rangle_E$. We call (\ref{BRmaster}) the Bloch-Redfield equation.
\subsubsection{Time-dependent Redfield tensor}
In the time-dependent version which we call as the time-dependent Bloch-Redfield (TDBR), we write the time evolution of the density matrix as
\begin{equation}
    \dot{\rho}_{mn}(t) = -i\omega_{mn}(t)\rho_{mn}(t) + \sum_{m',n'}\mathcal{R}_{mnm'n'}(t)\rho_{m'u'}(t)
    \label{RedfieldTL}
\end{equation}
where $\{\ket{m(t)}\}$ is the instantaneous eigenbasis of the driven Hamiltonian $\hat{H}_S(t)$ at time $t$ i.e., $\hat{H}_S(t)\ket{m(t)} = \varepsilon_m(t)\ket{m(t)}$ where $\varepsilon_m(t)$ are the instantaneous quasi-energies and the frequencies $\omega_{mn}(t) = \varepsilon_m(t) - \varepsilon_n(t)$. This approach incorporates the dressing due to the drive and thus in principle should be more accurate than considering the drive purely as part of the qubit-cavity Hamiltonian, but it is important to note that it is only reasonably valid for an adiabatic drive, meaning that the instantaneous energy gaps $\omega_{mn}(t)$ should rotate \textit{slowly} compared to the environment relaxation timescale $\tau_B\approx\omega_c^{-1}$. This means from the perspective of the environment, the system should be `frozen' in time. For the physical system under study, the periodicity of the instantaneous energy gaps is set by the drive frequency, which is always equal to the bare cavity frequency $\omega_r$; $\omega_c$ is taken as the largest frequency scale in the problem, and can be made arbitrarily large, in order to fulfill the requirements discussed above.  
\subsubsection{Secular approximation}
\begin{figure}
    \centering
    \includegraphics[width=0.8\columnwidth]{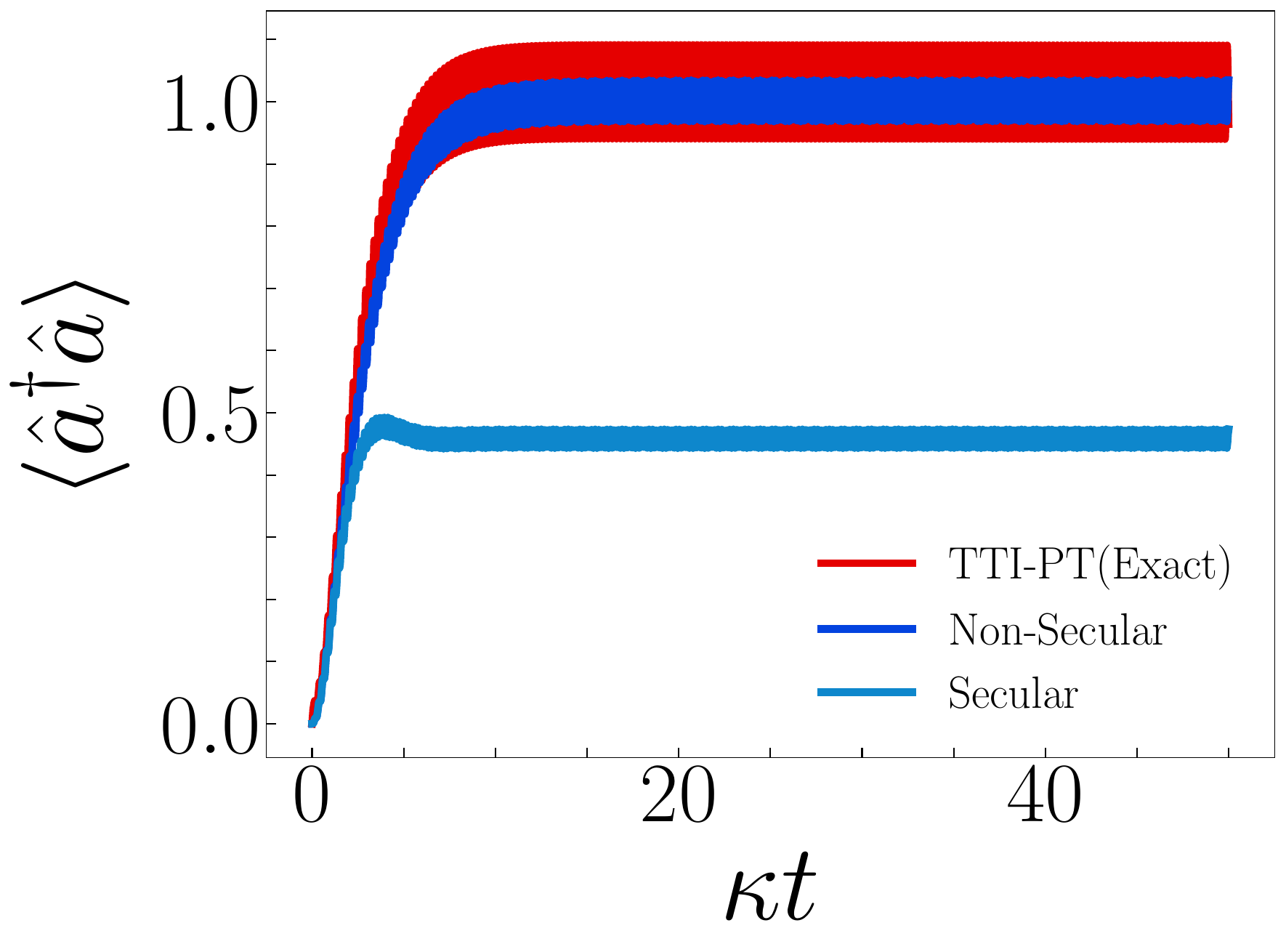}
    \caption{A figure to illustrate the effect of the secular approximation using a cavity driven on resonance as an example. The plots show photon number of the cavity as a function of the dimensionless parameter $\kappa t$ where $\kappa/2\pi = 0.5$ GHz is the single photon loss rate. The Bloch-Redfield results are benchmarked with the TTI-PT (Time Translation Invariant-Process Tensor) method \cite{cochin_processtensors_2026}, a non-perturbative technique that is numerically exact.}
    \label{fig:cav_bench}
\end{figure}

In equation (\ref{BRmaster}) instead of summing over all possible terms, we can impose a cutoff $\omega_f$ such that only terms that satisfy the relation $|\omega_{\mu\nu} - \omega_{\mu'\nu'}| < \omega_f$ are used to construct the dissipator. This is essentially built from the fact that we want to select a time scale $\tau^*$ such that 
\begin{equation}
    |\omega_{\mu\nu} - \omega_{\mu'\nu'}|^{-1} \ll \tau^* \ll \tau_R,
\end{equation}
so that while integrating equation (\ref{BRmaster}) in the interaction picture w.r.t. $\hat{H}_S+\hat{H}_E$, the fast oscillating quantities give zero contribution to the dynamics \cite{cattaneo_local_2019}. Here $\tau_R$ is set by the slowest relaxation timescale of the system. In our case this is set by the Purcell decay rate of the qubit. The full secular approximation is defined by the limit $\tau^* \rightarrow \infty$ or $|\omega_{\mu\nu} - \omega_{\mu'\nu'}| \rightarrow0$ and consequently, it involves removing all the terms from (\ref{BRmaster}) and (\ref{RedfieldTL}) for which $\omega_{\mu\nu} \neq \omega_{\mu'\nu'}$. We note that such a `secularization' cutoff can also be used in the time-dependent Redfield equation (\ref{RedfieldTL}). However, this approximation can be problematic in the situation where the system states have degenerate, or close-to degenerate, energy gaps, i.e. $\exists$ $\omega_{\mu\nu}$, $\omega_{\mu'\nu'}$ such that
\begin{equation}
     |\omega_{\mu\nu} - \omega_{\mu'\nu'}|^{-1} \gg \tau_R,
\end{equation}
for any frequency pairs $\omega_{\mu\nu}, \omega_{\mu'\nu'}$. Critically, for a driven harmonic resonator (decoupled from the qubit) all the eigenstate transitions are degenerate, we observe in Fig.(\ref{fig:cav_bench}) that enforcing the secular approximation in the Bloch-Redfield equation completely fails to accurately predict the steady state photon number. For these reasons, we only consider the full (non-secular) Bloch-Redfield equations in the results to follow.  
\section{Results}

\subsection{Purcell decay with undriven resonator}
We first test both the models for the undriven case i.e. Purcell decay. It is the relaxation of the qubit via a photon emission in the resonator. With $\ket{g}$ and $\ket{e}$ as the ground and excited states of the qubit respectively, it is defined as the $\ket{\overline{g0}}\leftarrow\ket{\overline{e0}}$ transition where $\ket{\overline{e0}}$ and $\ket{\overline{g0}}$ are the eigenstates of the Rabi Hamiltonian $\hat{H}_0$. For the system parameters we use typical values that have previously been used to investigate the `measurement problem' \cite{riva_dispersivereadout_2026, cochin_processtensors_2026}: $\omega_q/2\pi = 5.304$ GHz, $\omega_r/2\pi = 7.5$ GHz, $g/2\pi = 0.211$ GHz with the system initialized at $\ket{\overline{e0}}$ . The single photon loss rate $\kappa$ is also in angular units but we will avoid writing $\kappa/2\pi$ everytime to increase readability. We probe both the regimes $g \gg \kappa$ and $g \ll \kappa$, which are we refer to as the strong and weak  (qubit-cavity) coupling regimes, respectively \cite{krantz_quantum_2019}. The strong coupling regime is particularly relevant for circuit-QED experiments \cite{wallraff_strong_2004, schuster_2005} and has been extensively studied with the Lindblad model.

\begin{figure}[h]
  \centering
  \begin{subfigure}[b]{0.5\columnwidth}
    \centering
    \includegraphics[width=\linewidth]{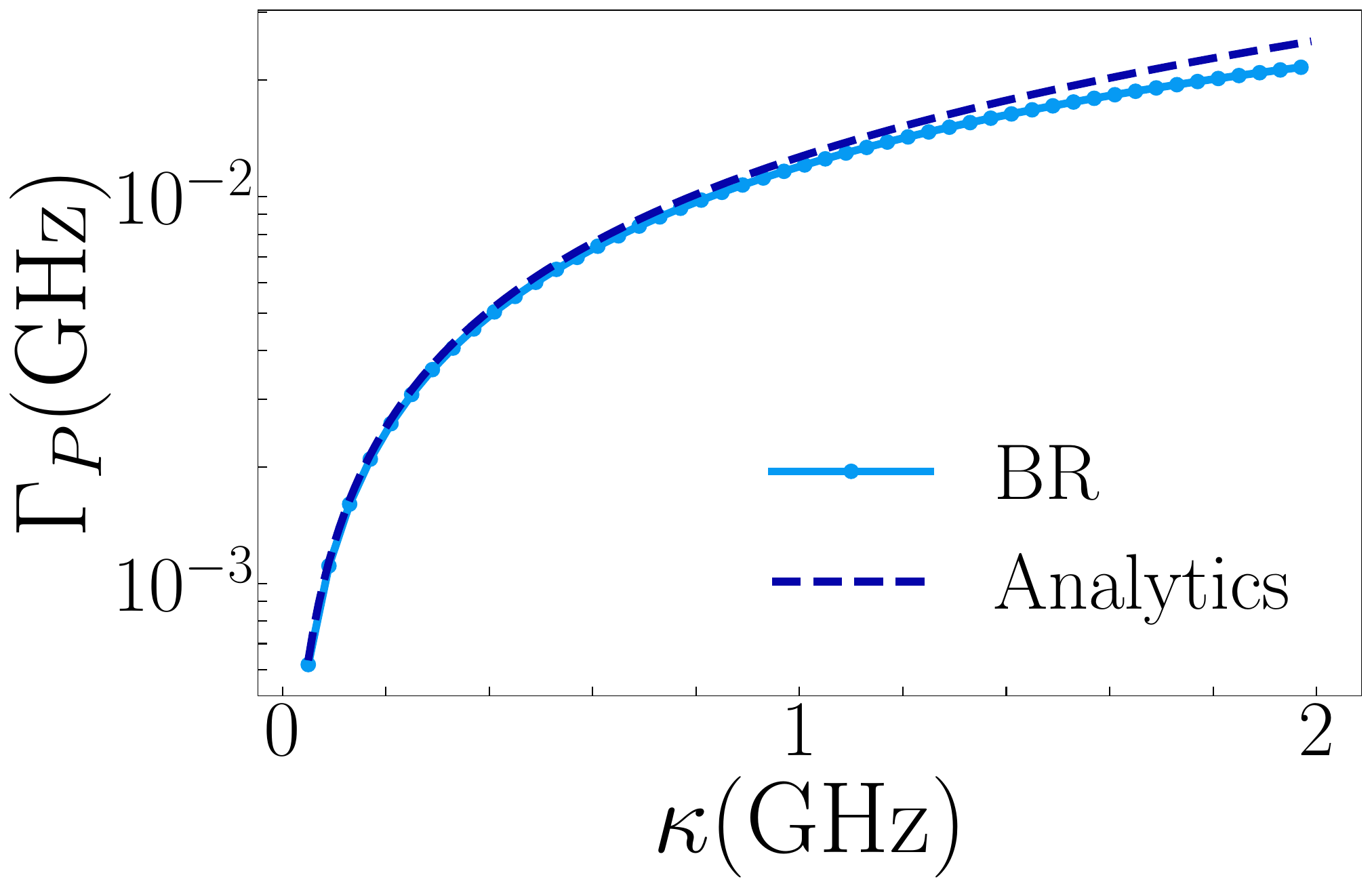}
    \caption{}
    \label{fig_purcell:subfig1}
  \end{subfigure}\hfill
  \begin{subfigure}[b]{0.47\columnwidth}
    \centering
    \includegraphics[width=\linewidth]{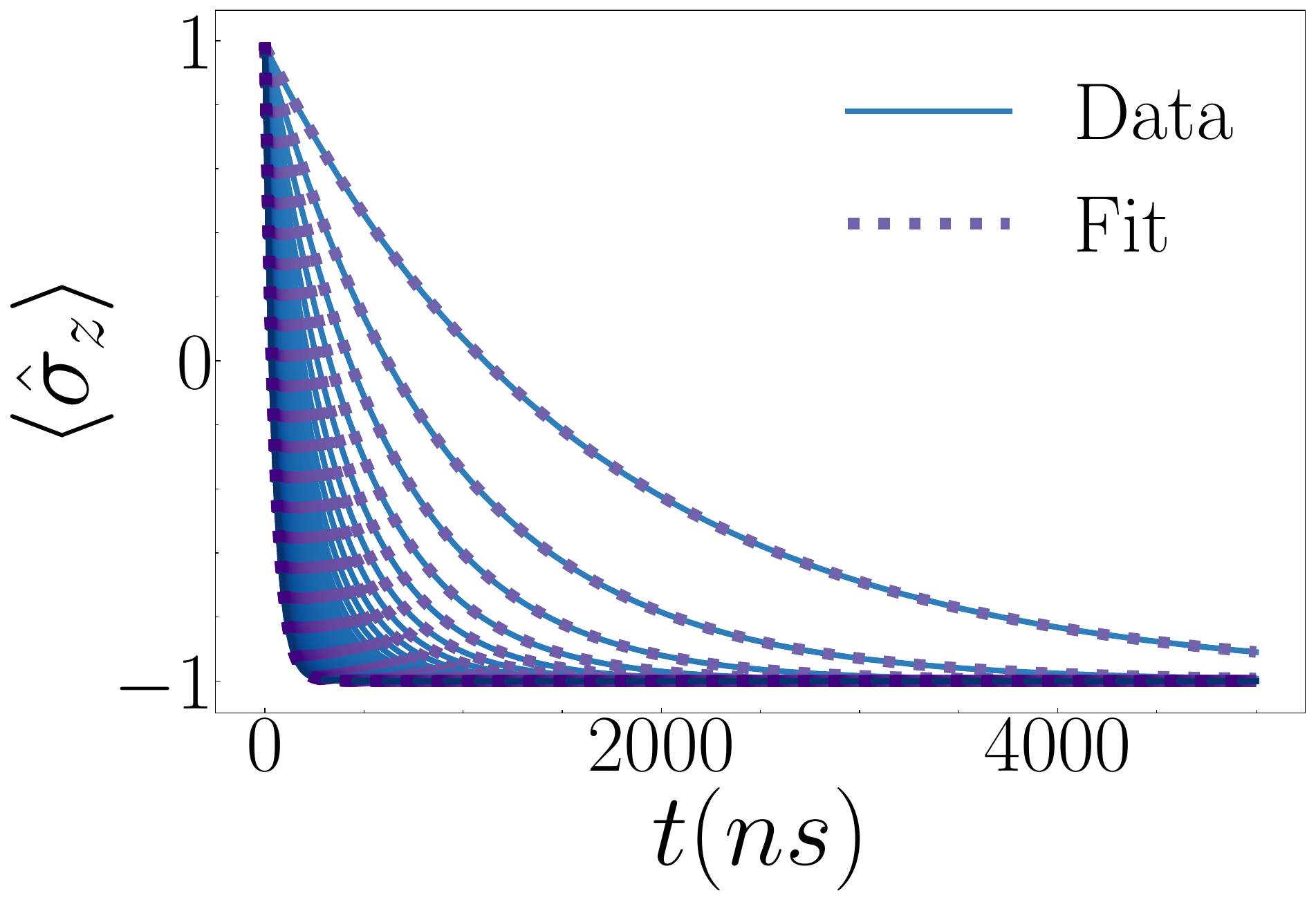}
    \caption{}
    \label{fig_purcell:subfig2}
  \end{subfigure}\hfill
  \begin{subfigure}[b]{0.80\columnwidth}
    \centering
    \includegraphics[width=\linewidth]{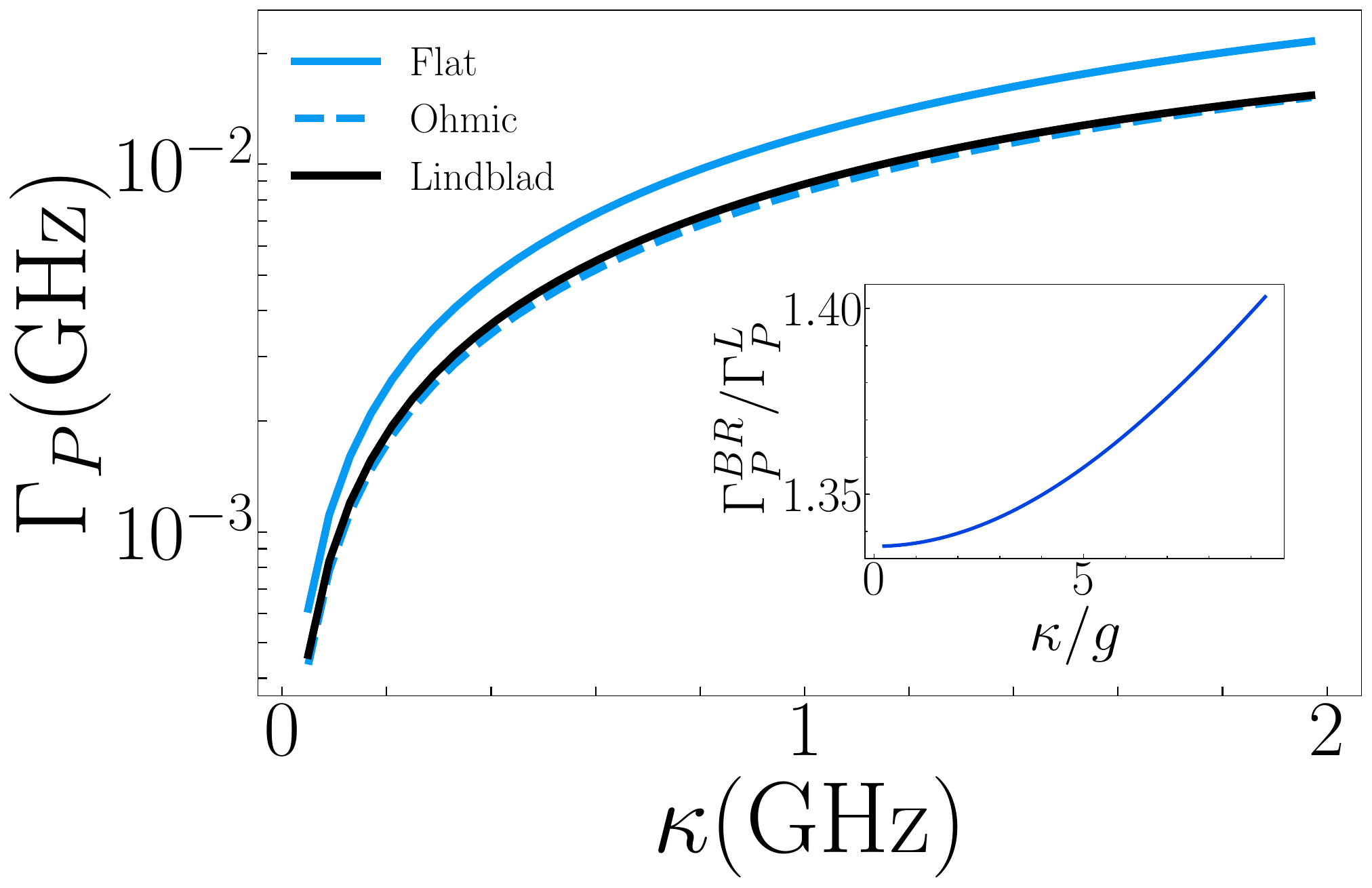}
    \caption{}
    \label{fig_purcell:subfig3}
  \end{subfigure}
  \caption{
  (a): Purcell rate $\Gamma_P$ vs. $\kappa$ behavior for the Bloch-Redfield simulations for the range $\kappa \in [0.05, 2.0]$. The spectral density is considered to be flat in this case such that $J(\omega) = J(\omega_r) = \kappa$. The dashed line is for the rates calculated from the expression (\ref{rateanalyt}). (b): $\langle\hat{\sigma}_z\rangle$ dynamics with increasing $\kappa$ and the dotted line shows the data fits used to calculate the Purcell rates. (c): Bloch-Redfield vs. Lindblad simulations. The black line represents the Lindblad, the solid blue line is for the Bloch-Redfield with a flat spectral density and the dashed blue line shows Bloch-Redfield rates with an Ohmic spectral density (\ref{sdActual}). The inset is the ratio $\Gamma^{BR}_P/\Gamma^L_{P}$ calculated from the simulations as a function of $\kappa/g$.}
  \label{purcell_fig}
\end{figure}

To treat both the models, Lindblad and Bloch-Redfield from an equal footing with respect to the noise spectrum, we first consider a flat (frequency independent) spectral density instead of the Ohmic form. The rates are calculated by fitting an exponential of the form
\begin{equation}
    f(t) = Ae^{-\gamma t} + B
\end{equation}
to the $\langle\hat{\sigma}_z\rangle$ dynamics. From Fig.(\ref{purcell_fig}) it is seen that on using a flat spectral density such that $J(\omega) \equiv const = \kappa$ for the Bloch-Redfield simulations, the results deviate from the Lindblad dynamics much more in comparison to using an Ohmic spectral density. This is because if we compare the Purcell decay rate for both the Lindblad and Bloch-Redfield, we get
\begin{equation}
    \frac{\Gamma^{BR}_P}{\Gamma^L_{P}} = \frac{J(\omega_{q})|\bra{\overline{g0}}\hat{X}\ket{\overline{e0}}|^2}{\kappa|\bra{\overline{g0}}\hat{a}\ket{\overline{e0}}|^2},
    \label{rates_compare}
\end{equation}
where $\hat{X} = \hat{a} + \hat{a}^\dagger$, is the displacement operator of the cavity. For the Rabi Hamiltonian, in the dispersive regime ($|g/\Delta|\ll1,\ \Delta = \omega_q - \omega_r$), the eigenstates $\ket{\overline{e0}}$ and $\ket{\overline{g0}}$ up-to first order in $|g/\Delta|$ are given by
\begin{subequations}\label{eq:egstates_rabi}
    \begin{align}
        \ket{\overline{e0}} &\simeq \ket{e0} + \frac{g}{\Delta}\ket{g1} + \mathcal{O}(|g/\Delta|^2), \label{eq:egstates_rabia}\\
        \ket{\overline{g0}} &\simeq\ket{g0} - \frac{g}{\Sigma}\ket{e1} + \mathcal{O}(|g/\Delta|^2).\label{eq:egstates_rabib}
    \end{align}
\end{subequations}
where $\Sigma = \omega_q + \omega_r$. That gives us 
\begin{equation}
    \Gamma_P^{BR} =J(\omega_{q})\left|\frac{g(\Sigma - \Delta)}{\Sigma\Delta}\right|^2=\kappa\left|\frac{2\omega_r g}{\omega_q^2 - \omega_r^2}\right|^2
    \label{rateanalyt}
\end{equation}
for $J(\omega_{q}) =\kappa$.
Hence the ratio (\ref{rates_compare}) for a Bloch-Redfield model with flat spectral density and Lindblad is
\begin{equation}
    \frac{\Gamma^{BR}_P}{\Gamma^L_{P}} = \frac{\kappa\left|\frac{g}{\Delta} - \frac{g}{\Sigma}\right|^2}{\kappa\left|\frac{g}{\Delta}\right|^2} \approx \left|\frac{2\omega_r}{\omega_r + \omega_q}\right|^2,
    \label{pcrate_cmp}
\end{equation}
which is approximately constant for the regime $\kappa \ll g$ \cite{muller_dissipativerabi}.

Thus a stronger Purcell rate for the Bloch-Redfield model can be traced to the fact that the Rabi ground state $\ket{\overline{g0}}$ contains a qubit-cavity excitation, in contrast to the Jaynes-Cummings ground state which is simply $\ket{g0}$. Consequently, the displacement operator $\hat{X}$ also couples the $\ket{e1} \leftarrow\ket{e0}$ components of the eigenstates, giving rise to the extra $-g/\Sigma$ term in the Purcell rate. In the strongly detuned case ($\omega_r\gg\omega_q$), this results in an four-fold increase in the decay rate. On the other hand in the Lindblad model, the single photon loss operator $\hat{a}$ is not able couple this transition and thus has a reduced decay rate in comparison to the Bloch-Redfield. It can also be deduced that the Lindblad model is always going to give the same decay rate irrespective of the undriven system Hamiltonian we use. Appendix A illustrates this effect by comparing the two models for the Purcell decay with two different system Hamiltonians $\hat{H}_0$ and $\hat{H}_{JC}$.

In the Ohmic case, the fall in the rates in comparison to the flat spectrum is because the spectral weight in the Ohmic spectrum at the qubit frequency is much weaker than the weight in the flat spectrum which is given by $\kappa$. This compensates for the rise in the Purcell rate due to the extra term in the ground state (\ref{eq:egstates_rabib}) and hence approaches the Lindblad rates.

However as we increase  $\kappa$ Fig.(\ref{fig_purcell:subfig1}), we see that the analytical values calculated from Eq.(\ref{rateanalyt}) slowly begin to deviate from the numerically fitted relaxation rates. Previous work \cite{sete_purcell_2014, blais_cavity_2004} suggests that the Purcell rate $\Gamma_p = (g/\Delta)^2\kappa$ derived from the Lindblad model is only valid for the strong coupling ($g \gg \kappa$) regime. In Fig.(\ref{fig_purcell:subfig3}) it can be seen that such an argument is also true for the Bloch-Redfield where the Purcell rate is given by (\ref{rateanalyt}). This effect is clearly highlighted in the inset of Fig.(\ref{fig_purcell:subfig3}), where we can observe that for higher values of $\kappa/g$, the ratio $\Gamma_P^{BR}/\Gamma_P^L$ has a non-linear dependence.
\subsection{Purcell decay with driven resonator}
\begin{figure}[h]
    \centering
    \begin{subfigure}[b]{0.85\columnwidth}
    \centering
    \includegraphics[width=\linewidth]{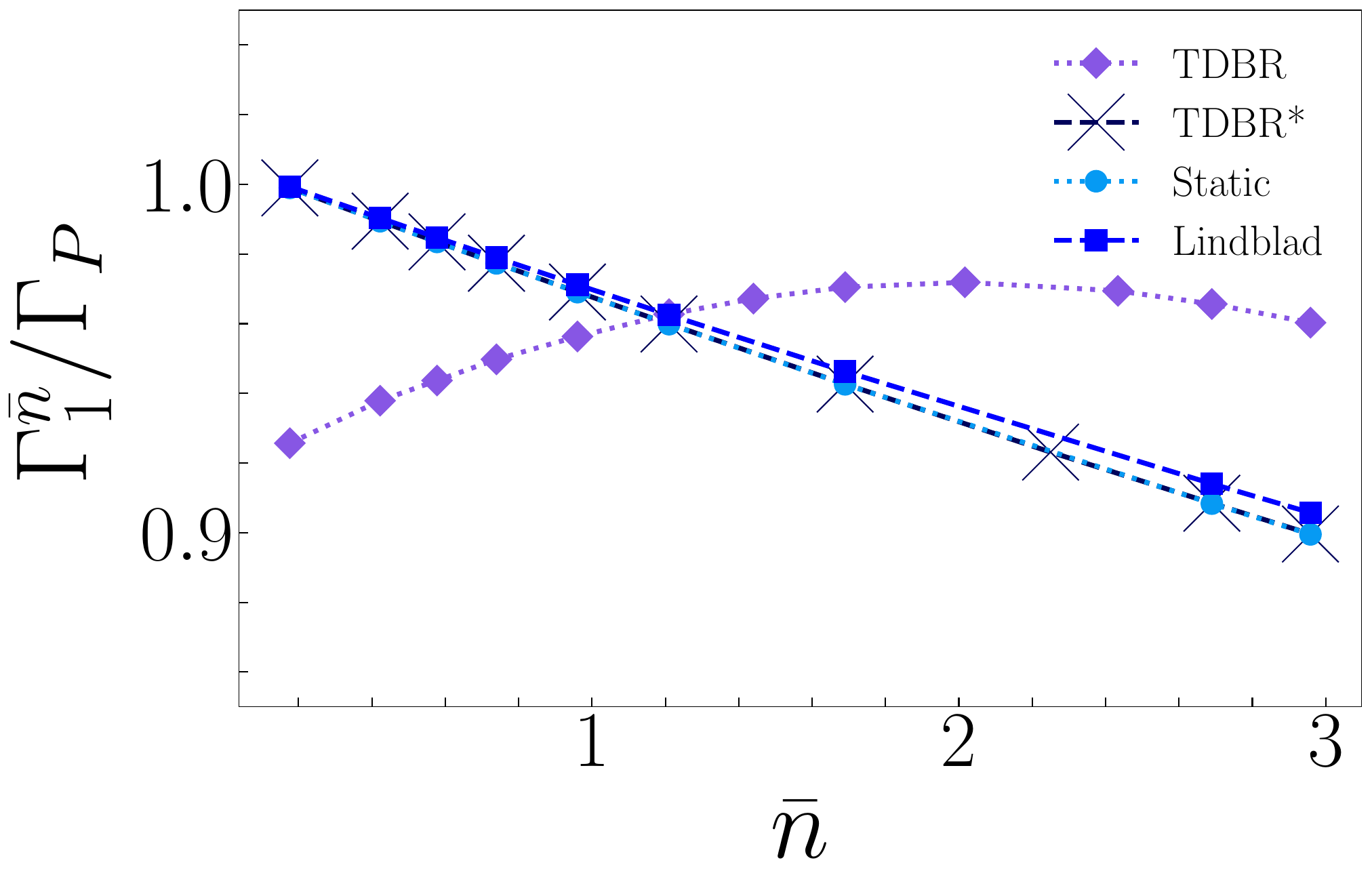}
    \caption{}
    \label{fig_driven:subfig1}
    \end{subfigure}\hfill
    \begin{subfigure}[b]{0.85\columnwidth}
    \centering
    \includegraphics[width=\linewidth]{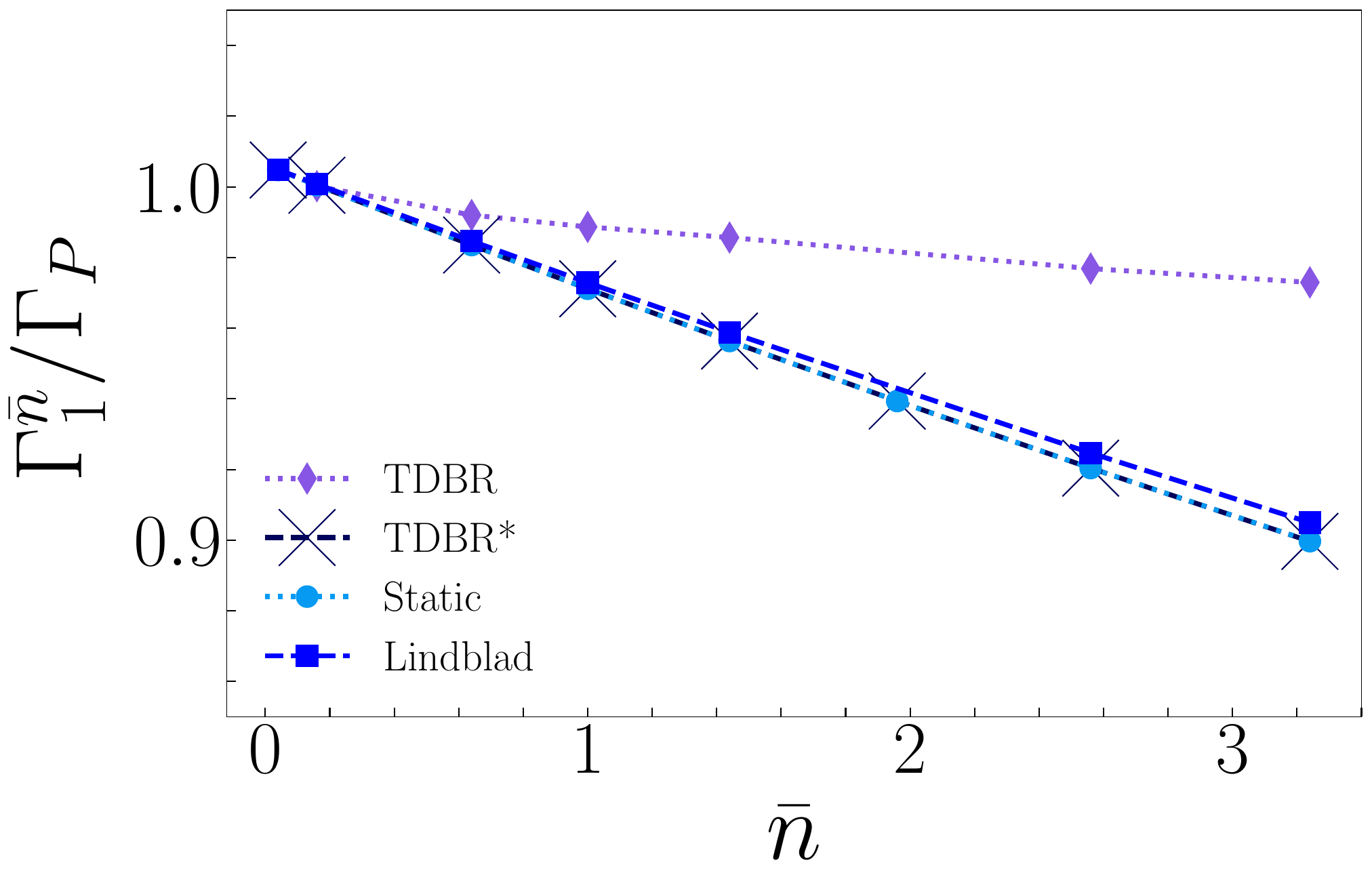}
    \caption{}
    \label{fig_driven:subfig2}
    \end{subfigure}
    \caption{(a): Relaxation rate $\Gamma^{\bar{n}}_1$ normalized by the Purcell rate $\Gamma_P$ for the respective methods as a function of mean photon number $\bar{n}$ in the cavity for $\kappa = 1.0$. Here the dotted purple curve show the $\Gamma_1(\bar{n})$ versus $\bar{n}$ behavior for the time-dependent Bloch-Redfield (TDBR) with an RWA drive of the form (\ref{rwadrive}). The TDBR* shows the same behavior for a cosine drive (\ref{drivenrwa}). The \textit{`Static'} curve is for the Bloch-Redfield model with a static dissipator. (b): $\Gamma^{\bar{n}}_1$ versus $\bar{n}$ for $\kappa = 0.1$.}
    \label{fig:g1nbar3methods}
\end{figure}

We now consider the driven system, with a drive on resonance ($\omega_d = \omega_r)$ applied the resonator and use the Ohmic spectral density (\ref{sdActual}) function to model the environment. First we consider the drive of the form
\begin{equation}
    \hat{H}_d(t) = \epsilon(\hat{a}e^{i\omega_d t} + \hat{a}^\dagger e^{-i\omega_dt})
    \label{rwadrive}
\end{equation}
with $\omega_d$ as the drive frequency. The counter rotating terms are dropped since the drive is on resonance with the resonator and we will call this the `RWA drive'. We again simulate the system for the same parameters as in the undriven case and the system initialized at $\ket{\overline{e0}}$ eigenstate. For the resonator photon loss rate, we consider $\kappa = 1.0$ and $\kappa = 0.1$ to study both the weak and strongly coupling regimes, respectively. From Fig.(\ref{fig:g1nbar3methods}), it is seen that on using the time-dependent prescription for constructing the Bloch-Redfield equation, the $\Gamma^{\bar{n}}_1$ vs. $\bar{n}$ behavior is completely different from the static Bloch-Redfield and the Lindblad results, initially \emph{increasing} with cavity population, and then decreasing at strong driving.

The non-monotonic behavior disappears once we consider a drive of the form
\begin{equation}
    \hat{H}_d(t) = \eta\cos(\omega_d t)(\hat{a} + \hat{a}^\dagger), 
    \label{drivenrwa}
\end{equation}
with $\eta = 2\epsilon$, as seen in Fig.(\ref{fig:g1nbar3methods}), where we observe the expected drive-induced suppression of the qubit's (Purcell) relaxation rate. 

This can be traced back to the time-dependent Bloch-Redfield formulation, where dissipation depends on the instantaneous matrix elements $|\bra{m(t)}\hat{X}\ket{n(t)}|^2$. A RWA drive removes counter-rotating terms, reshaping these matrix elements and producing different trends in $\Gamma^{\bar{n}}_1$ even at resonance. Restoring the full cosine drive that includes the counter-rotating terms cancels those spurious components and yields physically consistent, drive-induced suppression of the relaxation rate. In other words, over a period with the RWA drive, $\hat{H}_d(t)$ is never $0$ while with the cosine drive $\hat{H}_d(t)$ is $0$ every half a period. This difference is essentially responsible for the unphysical decay patterns that arise while using the time-dependent Bloch-Redfield (TDBR) method with an RWA drive. 

Thus we observe that even at resonance, depending on how we construct our dissipator, the nature of the drive can have non-negligible impact on the dynamics of the physical observables and so, from now on we will only work with a full drive without the rotating wave approximation.
Apart from that we also observe that as we increase the driving strength, there is a small quantitative difference on the suppression of the relaxation rates between the Bloch-Redfield and Lindblad models and the reason can be traced back to similar conclusions that we made in the undriven case where the differences arose due to the coupling operators that causes transitions between the eigenstates.

\begin{figure}[h]
    \centering
    \begin{subfigure}[b]{0.5\columnwidth}
    \centering
    \includegraphics[width=\linewidth]{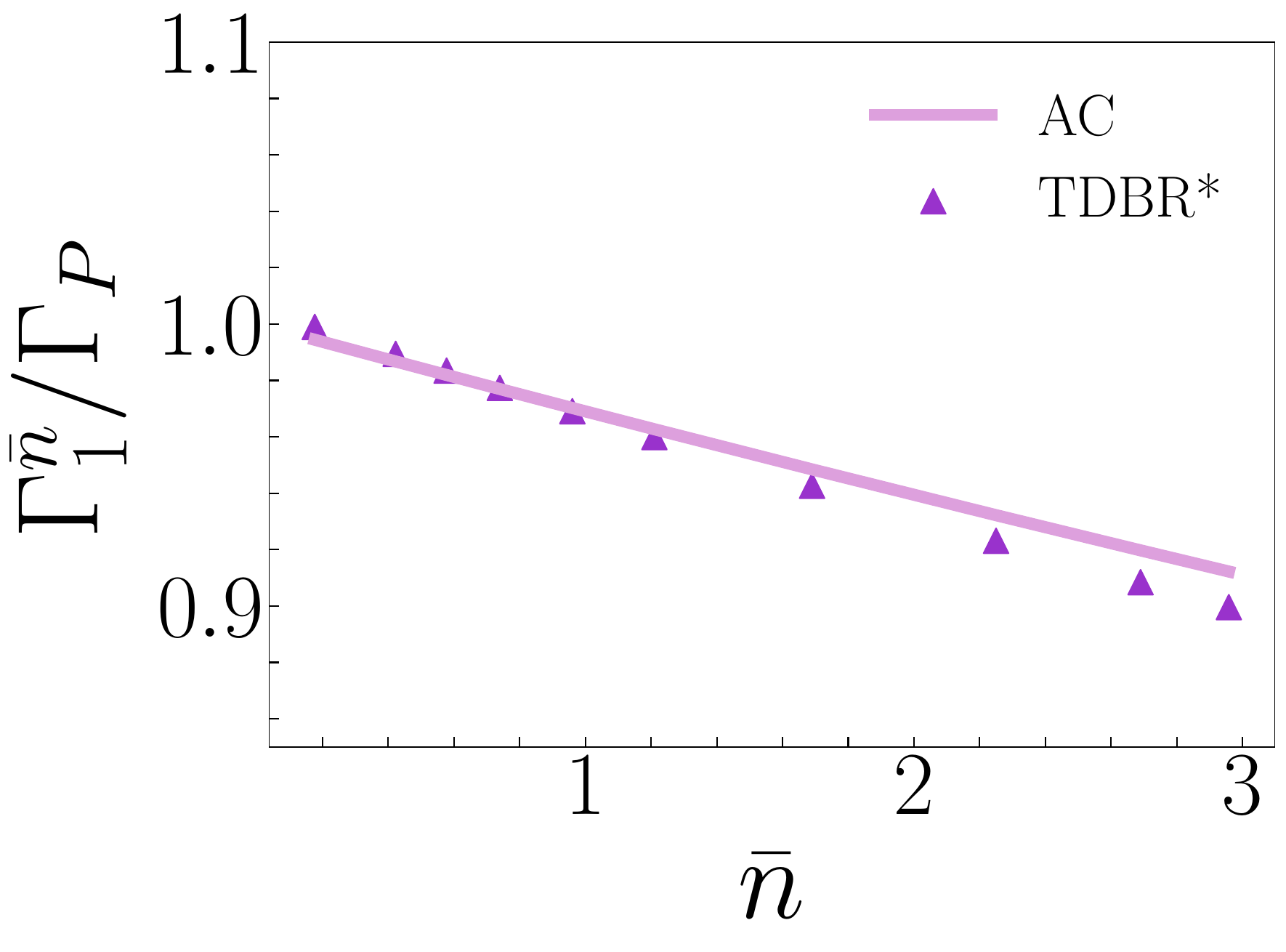}
    \caption{}
    \label{rates_driven:subfig1}
    \end{subfigure}\hfill
    \begin{subfigure}[b]{0.5\columnwidth}
    \centering
    \includegraphics[width=\linewidth]{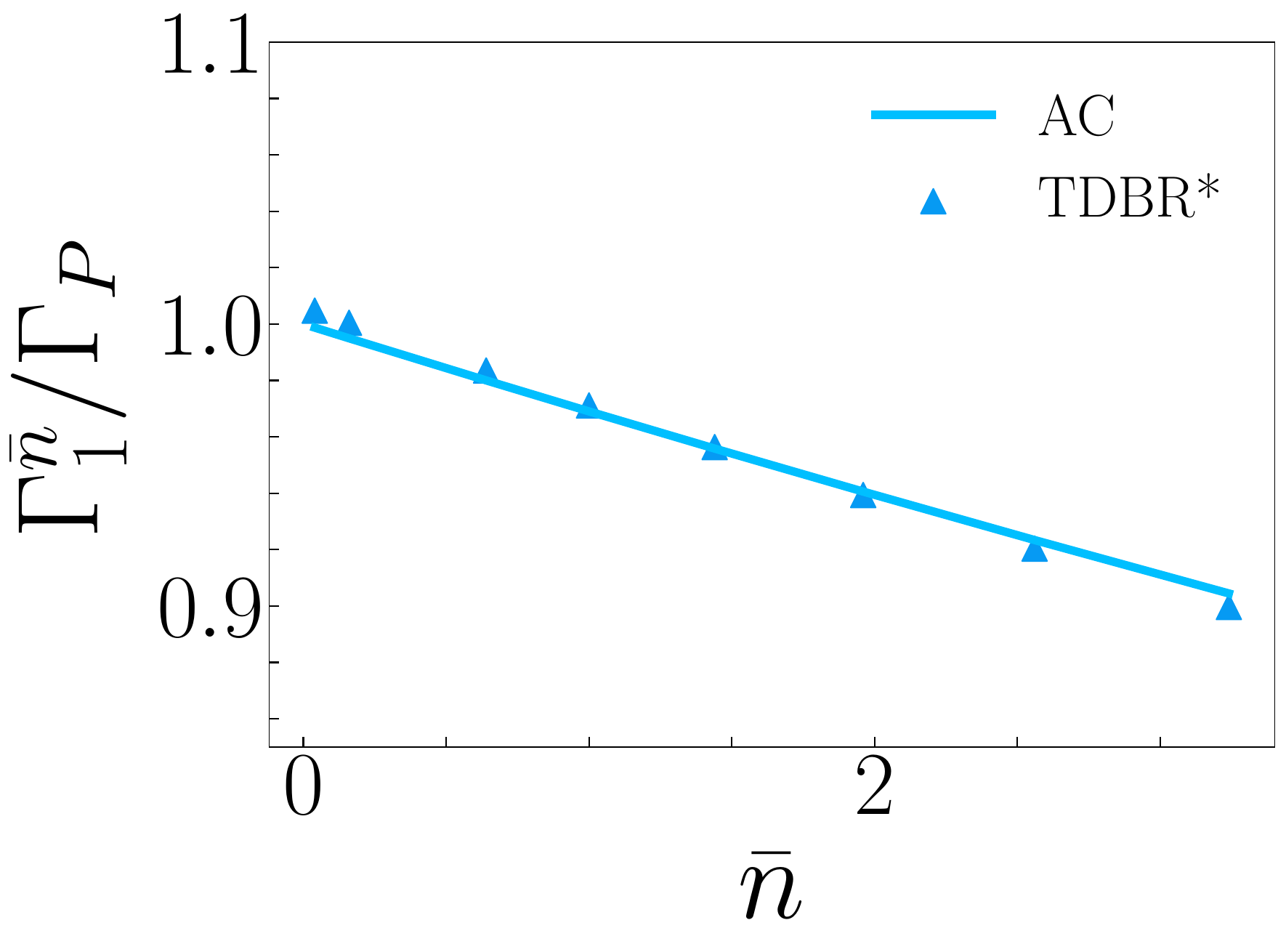}
    \caption{}
    \label{rates_driven:subfig2}
    \end{subfigure}
    \caption{(a): Comparison of the $\Gamma^{\bar{n}}_1$ versus $\bar{n}$ trend calculated from the numerical simulations and analytical expressions for $\kappa = 1.0$. The solid purple line shows the rates calculated from (\ref{ratesdriven}). (b): Similar comparison for $\kappa = 0.1$.}
    \label{fig:drivenrates_an}
\end{figure}

A physical reason behind the suppression of the relaxation rates is the AC-stark shift that modifies the bare qubit frequency $\omega_q$ by $\omega^{\bar{n}}_q = \omega_q + 2\chi\bar{n}$ \cite{thorbeck_2024}, where the dispersive shift $\chi$ is defined as $\chi = g^2(\Delta^{-1} + \Sigma^{-1})$ \cite{beaudoin_dissipation_2011}, and causes the qubit transition frequency to decrease. If the environment spectral function contains significant frequency dependence, as in the Ohmic case, then this suppression of the qubit transition frequency would modify the decay rate.  From (\ref{rateanalyt}) if we express the rate as a function of $\bar{n}$, we get
\begin{equation}
    \Gamma^{\bar{n}}_1 = J(\omega^{\bar{n}}_q)\left|\frac{2\omega_r g}{(\omega^{\bar{n}}_q)^2 - \omega_r^2}\right|^2.
    \label{ratesdriven}
\end{equation}
From Fig.(\ref{fig:drivenrates_an}) we see good agreement between the stark-shifted rates and the actual rates calculated from the numerical simulations. The mean error for $\kappa=1.0$ GHz ($\kappa = 0.1$ GHz) was found to be $\approx 0.35\%$ ($\approx 0.11\%$).
\subsection{Purcell decay with driven resonator and a bandpass filter}
\label{purcell_filt_sec}
To demonstrate the advantage of the Bloch-Redfield method over the Lindblad model with an emphasis on incorporating arbitrary environment spectral function, we consider the case of a Purcell filter that is commonly used in the circuit QED experiments \cite{reed_fast_2010, jeffrey_2014} to protect the qubit from the readout line noise. We will specifically consider a bandpass filter \cite{sete_bp, bronn_broadbandfilter_2015, yan_bandpasspurcell_2023}, that allows Purcell rate reduction for a wide range of qubit frequencies that are sufficiently detuned from the filter frequency. Additionally it also allows multiplexed readout as several qubits can be put into the stop band with the same filter \cite{bakr_reentrantfilter_2025, xiong_tunablepurcell_2026}. The filter can be modeled into the environment by modifying the spectral density as 
\begin{equation}
    J_{eff}(\omega) = J(\omega)\mathcal{F}_{bp}(\omega),
    \label{specBP}
\end{equation}
where $J(\omega)$ is the main readout line with the Ohmic spectral density and $\mathcal{F}_{bp}(\omega)$ is a function that describes the filter. In this work we choose $\mathcal{F}_{bp}(\omega)$ as 
\begin{equation}
    \mathcal{F}_{bp}(\omega) = \frac{1}{1+\left(\frac{\omega - \omega_f}{\gamma_f}\right)^2},
\end{equation}
where $\omega_f$ is the filter frequency and $\gamma_f$ is the parameter that controls the bandwidth of the filter. We have centred our filter at the cavity frequency $\omega_r$ and we choose two bandwidths, $\gamma^1_f = 1.5$ GHz and $\gamma^{2}_f = 1.0$ GHz. The single photon loss rate $\kappa$ is fixed at $0.1$ GHz.
\begin{figure}[h]
    \centering
    \begin{subfigure}[b]{0.85\columnwidth}
    \centering
    \includegraphics[width=\linewidth]{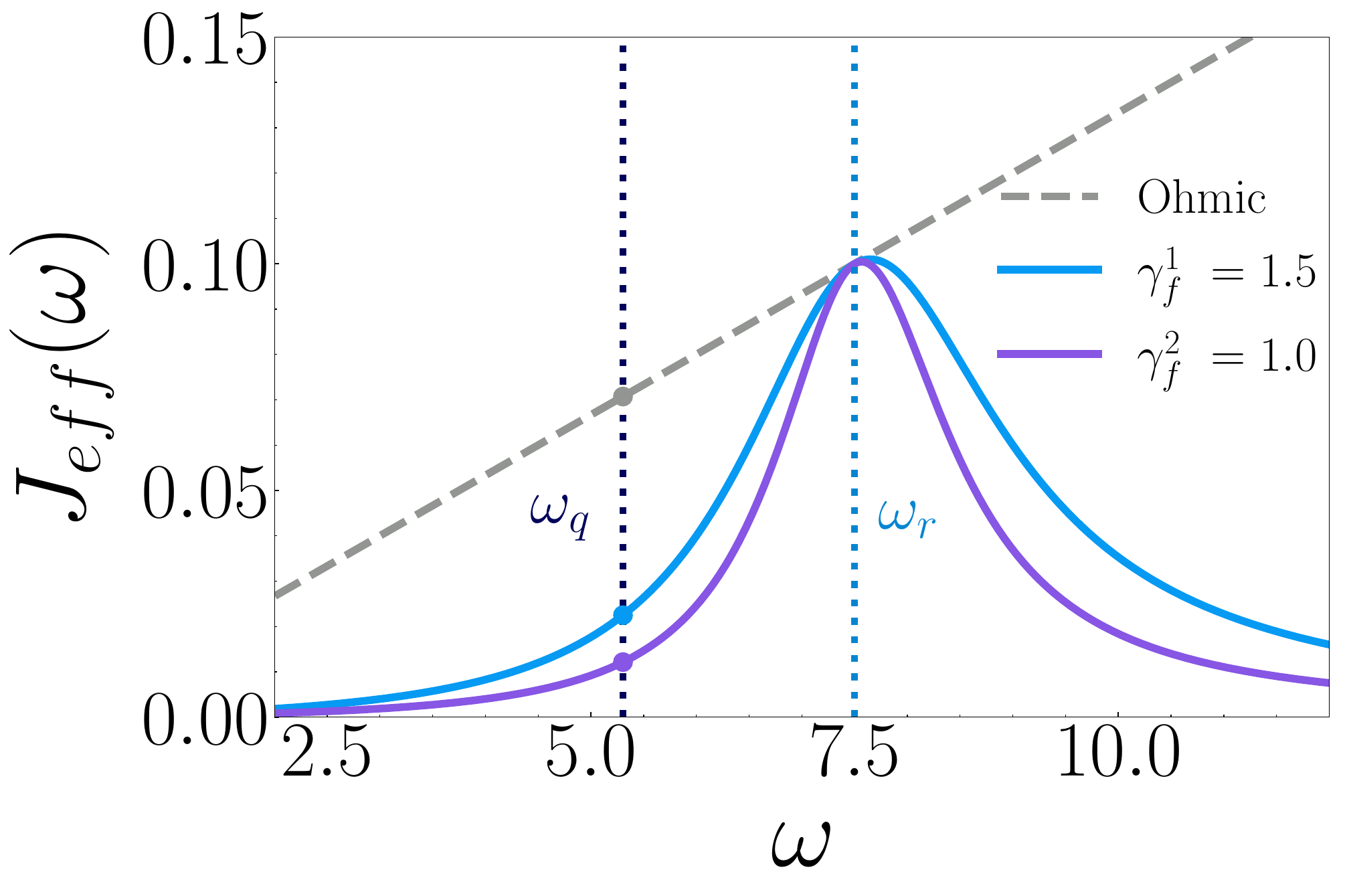}
    \caption{}
    \label{bpf_purcell:a}
    \end{subfigure}\hfill
    \begin{subfigure}[b]{0.85\columnwidth}
    \centering
    \includegraphics[width=\linewidth]{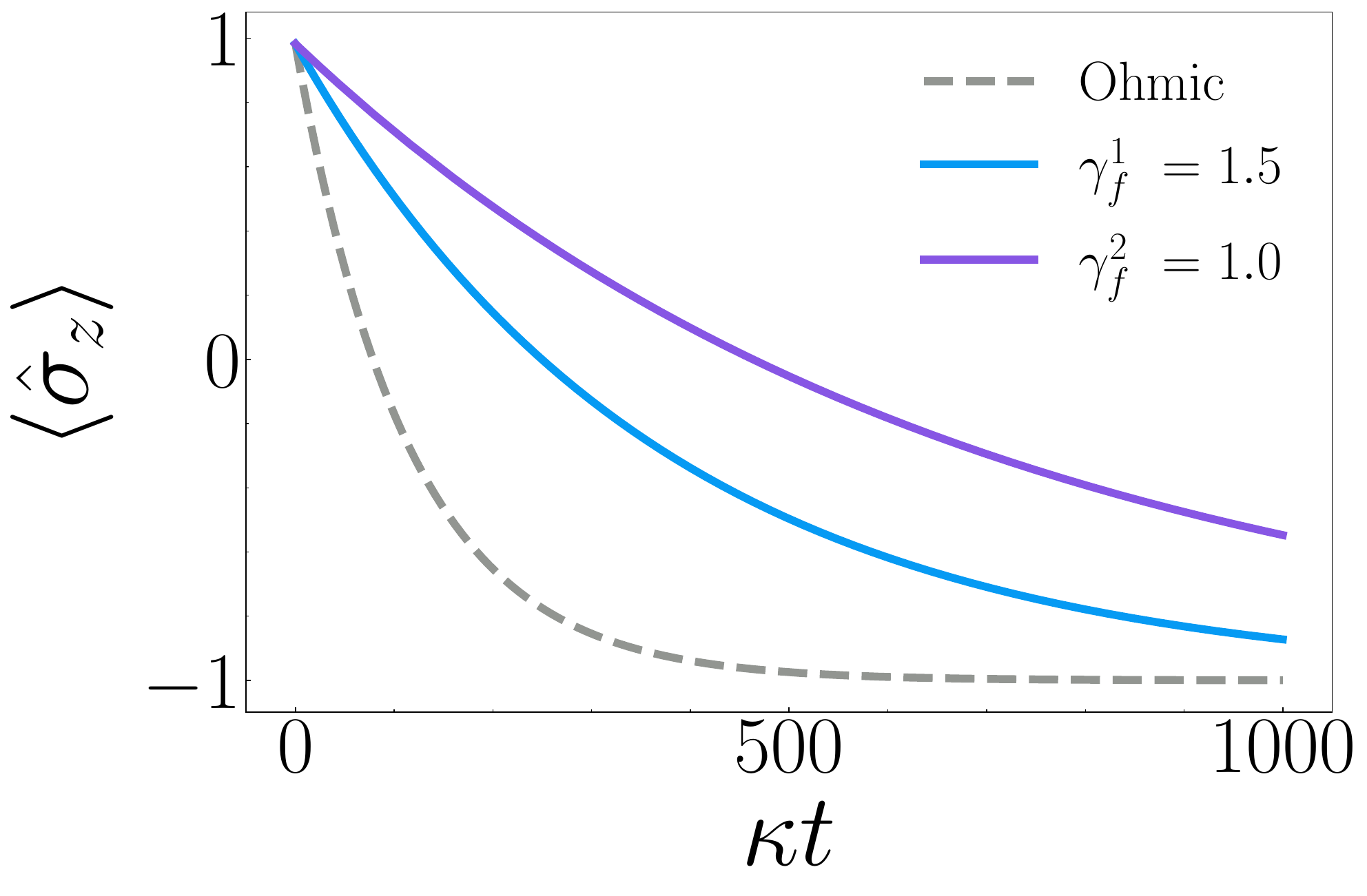}
    \caption{}
    \label{bpf_purcell:b}
    \end{subfigure}
    \caption{(a): Effective spectral density after introducing the bandpass filter. The two curves correspond to filters with bandwidths $\gamma^1_f = 1.5$ GHz and $\gamma^2_f = 1.0$ GHz respectively. The dashed grey line is a reference to the transmission line without the filter which is purely Ohmic. (b): $\langle\hat{\sigma}_z\rangle$ versus $\kappa t$ for the Ohmic, $\gamma^1_f$ and $\gamma^2_f$ cases respectively at zero drive. From the data, the factor $\mathcal{G}$ was found to be $3.175$($5.895$) for $\gamma^1_f$($\gamma^2_f$).}
    \label{fig:bpf_purcell}
\end{figure}
First we look at the undriven case. In Fig.(\ref{bpf_purcell:b}) we see that 
qubit relaxation rate is slowed once the bandpass filter is introduced into the environment. We denote the zero drive rates with $J_{eff}(\omega)$ as $\widetilde{\Gamma}_0$ and with the Ohmic spectral density as $\Gamma_0$, to avoid confusion, and define a ratio 
\begin{subequations}\label{bpfgain}
    \begin{align}
        \mathcal{G} &= \frac{\Gamma_0}{\widetilde{\Gamma}_0}=\frac{J(\omega_q)}{J_{eff}{(\omega_q)}},\label{bpfgain:a}\\
        \mathcal{G}&=1 + \left(\frac{\Delta}{\gamma_f}\right)^2\label{bpfgain:b},
    \end{align}
\end{subequations}
to determine the $T_1$ enhancement of the qubit i.e. how much the bandpass filter is suppressing the noise at the qubit frequency. For $\gamma^1_f = 1.5$ GHz and $\gamma^{2}_f = 1.0$ GHz, this factor is approximately $3.14$ and $5.82$ respectively. This means the modified qubit relaxation rate is $3.14$$(5.82)$ times slower than the Ohmic Purcell rate i.e. the zero drive rate for a filter with bandwidth $1.5$ GHz ($1.0$ GHz) with $\kappa$ fixed at $0.1$ GHz.

\begin{figure}[h]
    \centering
    \includegraphics[width=0.85\columnwidth]{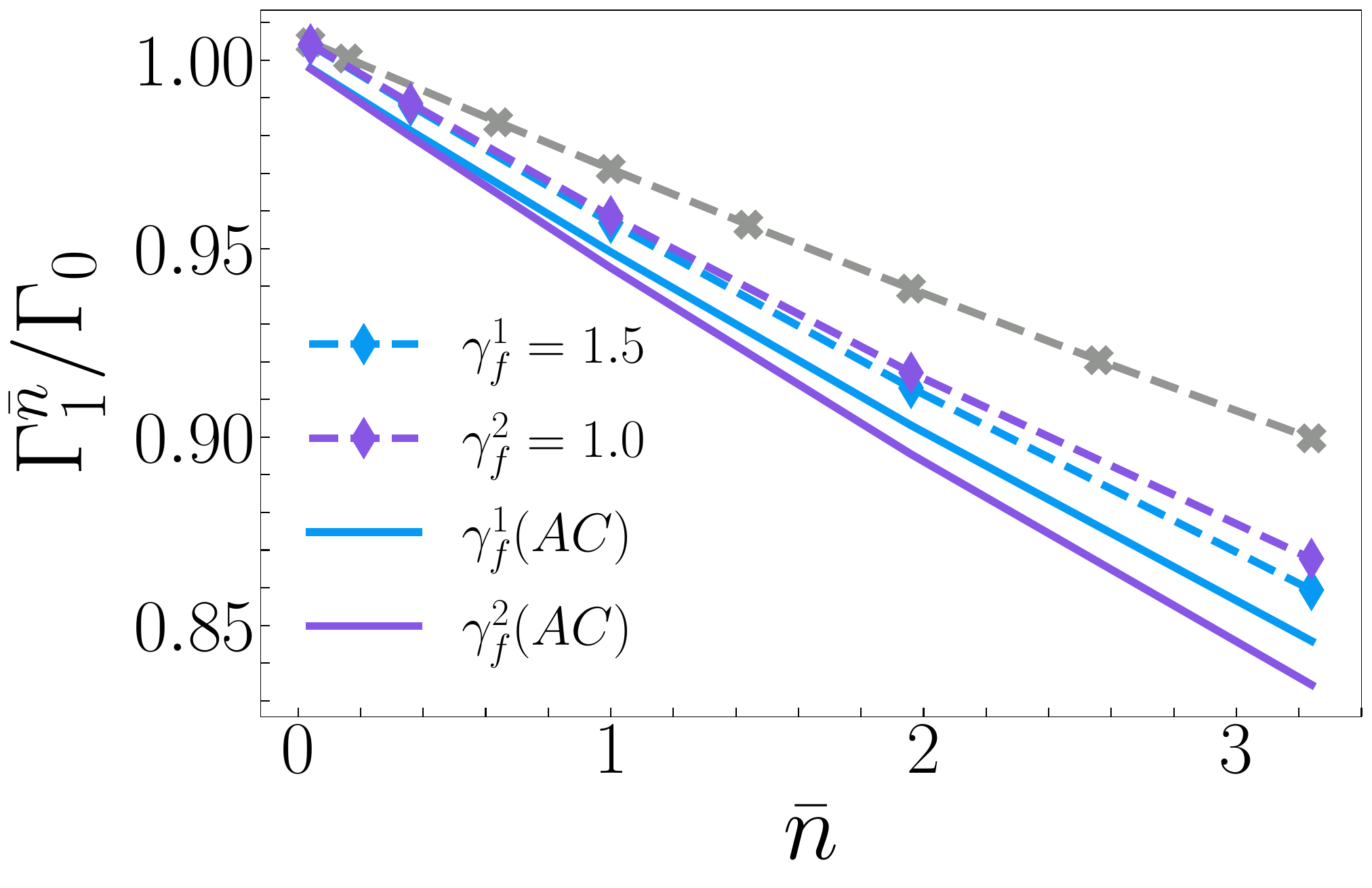}
    \caption{Drive induced decay rate $\Gamma^{\bar{n}}_1$ scaled by their respective zero drive rates for the Ohmic, $\gamma^1_f$ and $\gamma^2_f$. The dashed grey, blue and purple lines show the Ohmic, $\gamma_f^1$ and $\gamma_f^2$ rates respectively calculated from the simulations. The solid blue and purple lines show the rates calculated from (\ref{rateanalyt}) with the effective spectral density function for the $\gamma^1_f$ and $\gamma^2_f$ filters respectively.}
    \label{fig:g1nbbpf}
\end{figure}

In the driven case (Fig.\ref{fig:g1nbbpf}), one can observe that once the bandpass filter is introduced, the drive induced suppression is stronger in comparison to the Ohmic spectral density. The small quantitative mismatch between the AC stark shifted rates calculated from (\ref{ratesdriven}) and the actual rates calculated from the simulations, arises due to the fact that the dispersive shift $\chi$ for the Rabi Hamiltonian (\ref{modeleq: b}) is approximate up-to second order in $g$ . The difference is more pronounced in the bandpass filter case in comparison to the Ohmic as the environment spectrum is highly non-linear and small frequency errors can get amplified into visible rate errors. For these simulations we have only used the time-dependent Bloch-Redfield with the full drive.

\section{Conclusion}
We have presented a systematic comparison of Lindblad and Bloch-Redfield master equation approaches to the driven-dissipative dynamics of the dispersive readout circuit. In the undriven case, we showed that the two approaches yield quantitatively different Purcell decay rates, tracing the discrepancy to the counter-rotating terms in the Rabi Hamiltonian that are captured by the displacement operator in the Bloch-Redfield dissipator but not by the single-photon loss operator of the Lindblad equation. In the driven case, we demonstrated that the rotating-wave approximation applied to the drive Hamiltonian produces unphysical relaxation trends within the time-dependent Bloch-Redfield formulation, while retaining the full cosine drive restores physically consistent, drive-induced suppression of the Purcell rate. Finally, we showed that structured electromagnetic environments, including Purcell filters, can be incorporated directly through the spectral density function, avoiding the need to include filters as additional harmonic modes coupled to the system.

Several problems might be tackled naturally using this framework. Incorporating a multi-level weakly anharmonic model of the transmon qubit would allow the study of drive-induced transitions outside the computational basis. In this setting, the frequency selectivity of the Bloch-Redfield dissipator becomes particularly important: the anharmonicity of the transmon spectrum means that transitions between different levels occur at distinct frequencies, each coupling differently to the transmission-line environment, an effect that is entirely absent in the single-photon loss approximation.  The efficiency of the Bloch-Redfield approach in handling structured spectral densities makes it particularly well-suited for filter design and optimization at modest computational cost. Including finite-temperature effects would further allow the formalism to describe thermally activated excitation and detailed-balance corrections, extending its applicability to realistic operating conditions.
\begin{acknowledgments}
AWC acknowledges support from ANR project Radpolimer ANR-22-CE30-0033. PG acknowledges support from Paris Center of Quantum Technologies (QuanTEdu France). EC acknowledges funding from ENS Lyon (CDSN). AR acknowledges support from ANR project MecaFlux (ANR-21-CE47-0011).
\end{acknowledgments}
\section*{Data Availability}
The data that support the findings of this study are available from the corresponding author upon reasonable request.
\appendix
\section{Purcell Decay}
\begin{figure}[H]
    \centering
    \begin{subfigure}[b]{0.5\columnwidth}
    \centering
    \includegraphics[width=\linewidth]{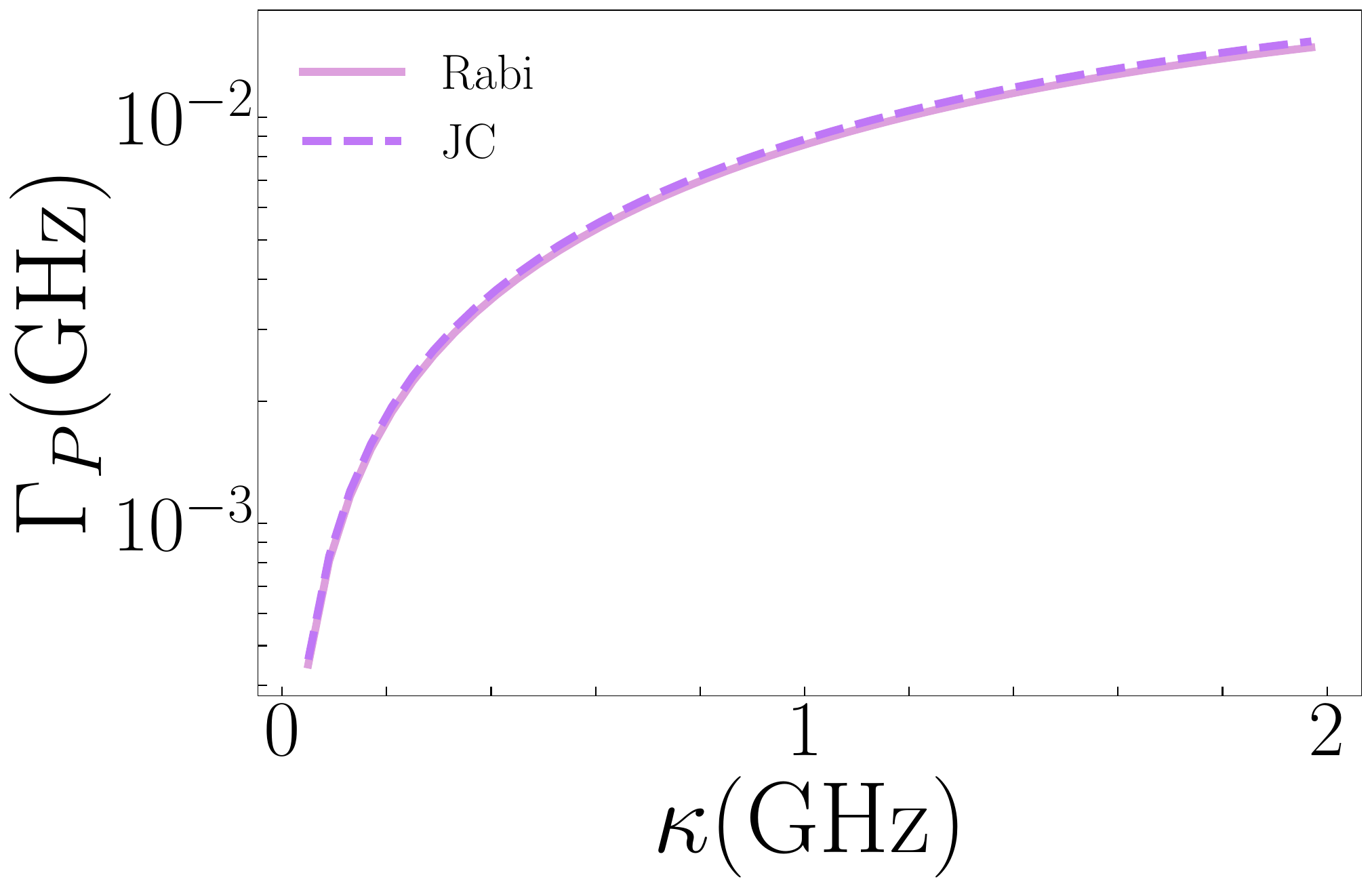}
    \caption{}
    \label{prtest:a}
    \end{subfigure}\hfill
    \begin{subfigure}[b]{0.5\columnwidth}
    \centering
    \includegraphics[width=\linewidth]{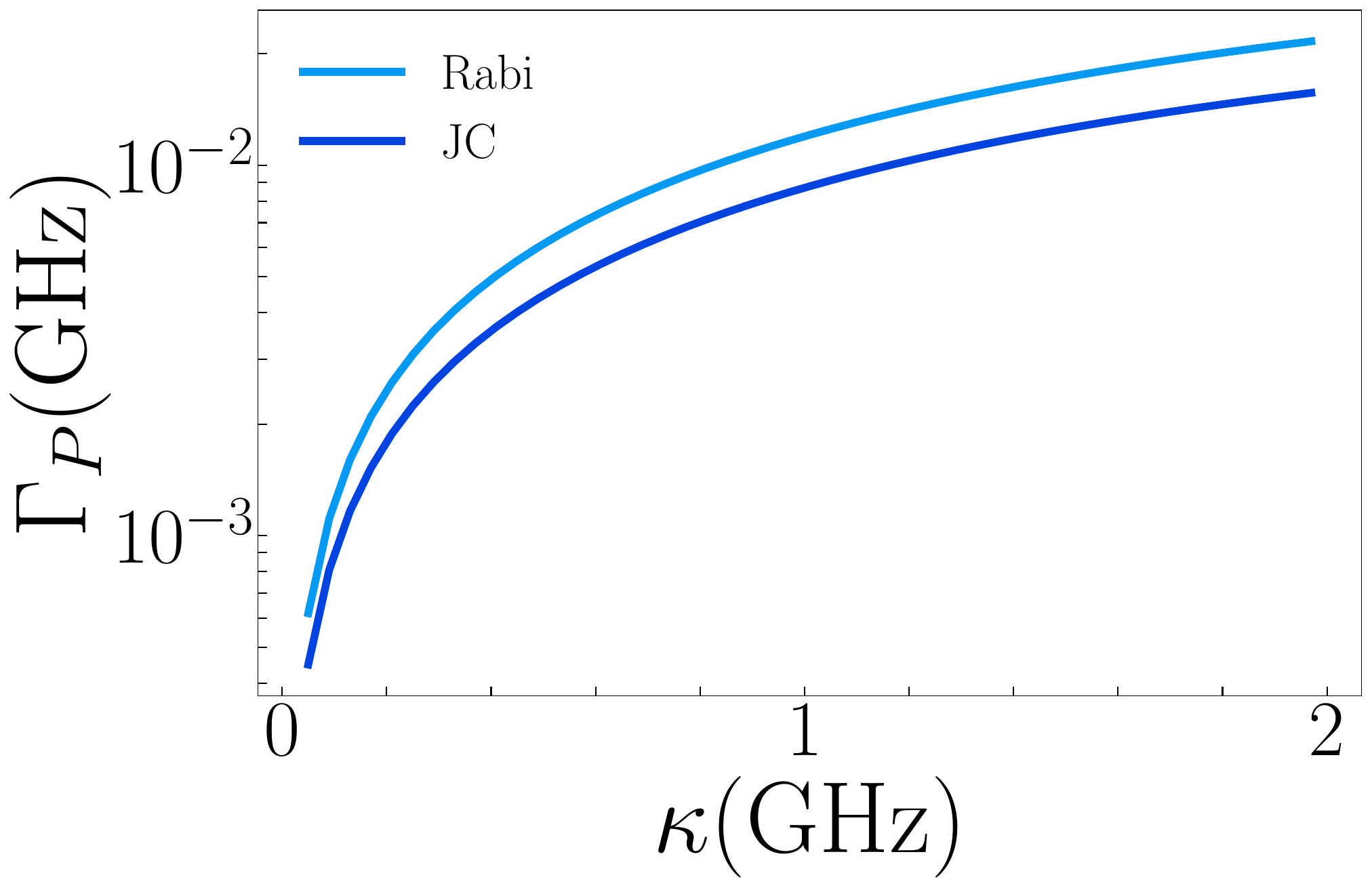}
    \caption{}
    \label{prtest:b}
    \end{subfigure}
    \caption{(a) shows the comparison of the Purcell decay with the Lindblad model (\ref{eq:Lindblad}) for $H_0$ and $H_{JC}$. Panel (b) is for the same comparison with the Bloch-Redfield model.}
    \label{prate_brl_rjc}
\end{figure}

Fig.(\ref{prate_brl_rjc}) shows a comparison of the Purcell decay rates with Bloch-Redfield and Lindblad for the flat spectral density and two different system Hamiltonians namely, $\hat{H}_0$ which is the Rabi Hamiltonian (\ref{modeleq: b}) and $\hat{H}_{JC}$ (\ref{eqjc}), the Jaynes-Cummings Hamiltonian.
\vspace{2mm}

\section{Bandpass Filter}
Fig.(\ref{fig:placeholder}) shows the $\langle\sigma_z\rangle$ dynamics for the two bandpass filters with bandwidth $\gamma^1_f$ and $\gamma^2_f$.
\begin{figure}[H]
    \centering
    \includegraphics[width=0.95\columnwidth]{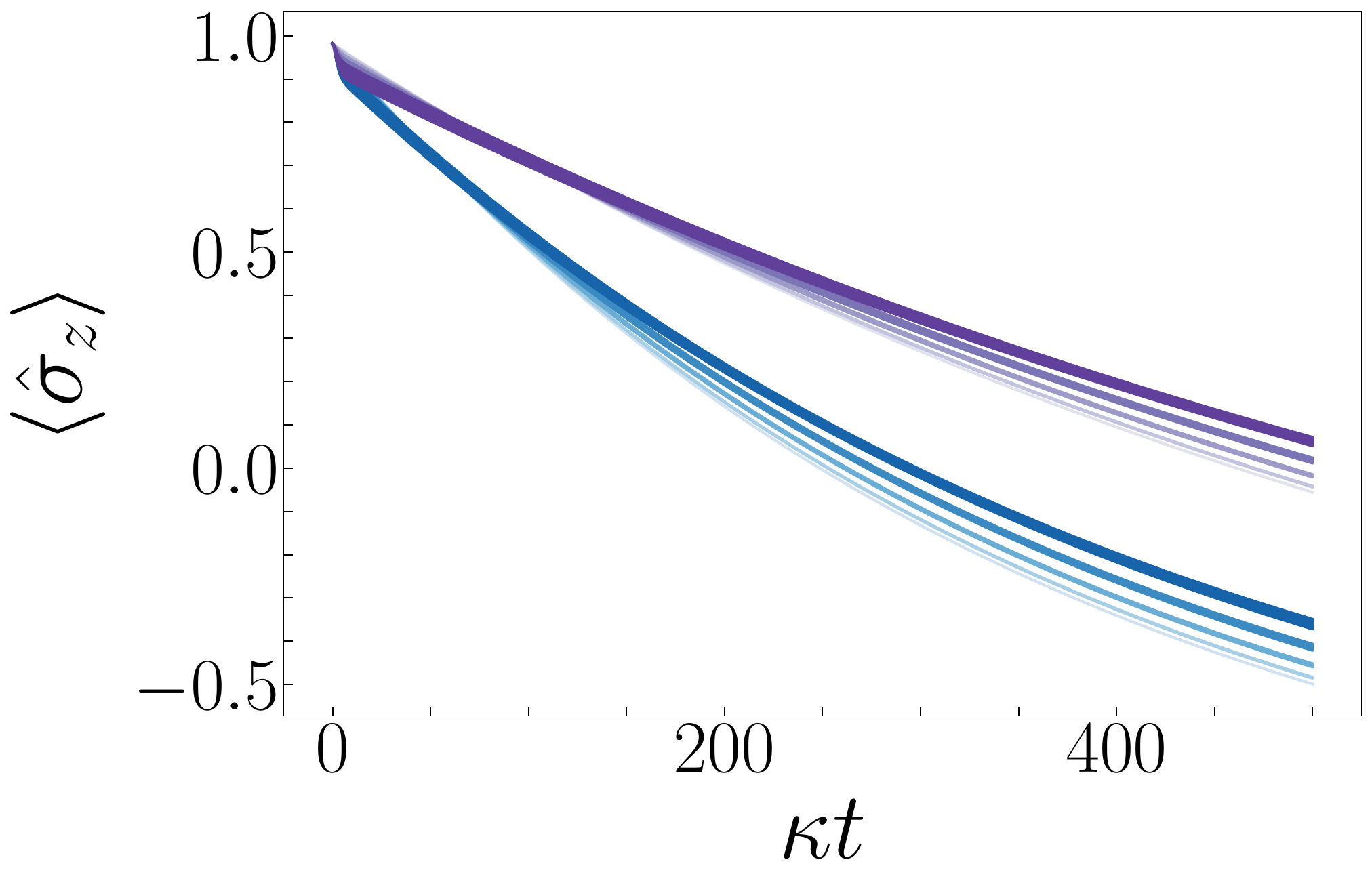}
    \caption{$\langle\hat{\sigma}_z\rangle$ vs. $t$ for the bandpass filters. The blue(purple) plots are for the filter with bandwidth $\gamma^1_f(\gamma^2_f)$. The darker colors are for the stronger drives and the lighter colors are for the weaker drives.}
    \label{fig:placeholder}
\end{figure}

\nocite{*}

\bibliography{main}

\end{document}